\documentclass[prc,twocolumn]{revtex4}
\usepackage{amsmath,amssymb,epsfig,bm,color,mathrsfs}
\newcommand{\be}{\begin{equation}}
\newcommand{\ee}{\end{equation}}
\newcommand{\bea}{\begin{eqnarray}}
\newcommand{\eea}{\end{eqnarray}}

\newcommand{\anu}{\bar\nu}

\newcommand{\ep}{\epsilon}

\newcommand{\Img}{{\rm Im}}
\newcommand{\vecq}{\bm q}
\newcommand{\vecv}{\bm v}
\newcommand{\vecp}{\bm p}
\newcommand{\veck}{\bm k}

\newcommand{\dF}{F^{+}}

\definecolor{red}{rgb}{0.8,0,0}
\definecolor{RED}{rgb}{0.8,0,0}
\definecolor{violet}{rgb}{0.4,0,0.4}
\definecolor{green}{rgb}{0,0.5,0.0}
\definecolor{GREEN}{rgb}{0,0.5,0.0}
\definecolor{navy}{rgb}{0.0,0.0,0.6}
\definecolor{orange}{rgb}{0.8,0.2,0.0}
\definecolor{blue}{rgb}{0.3,0.0,0.8}

\begin{document}
\title{
   Vertex renormalization of weak interactions in compact stars:\\
beyond leading order
}
\author{Armen Sedrakian\footnote{also Department of Physics, Yerevan
    State University, Armenia}}
\address{
Institute for Theoretical Physics, 
J.~W.~Goethe-University, \\D-60438  Frankfurt-Main, Germany
}

\begin{abstract}               

  Neutrino emission rate from baryonic matter in neutron stars via
  weak neutral vector interaction is computed up to order $O(v_F^6)$,
  where $v_F$ is the Fermi velocity in units of speed of light. The
  vector current polarization tensors are evaluated with full vertices
  which include resummed series in the particle-hole channel.  The
  neutrino emissivity is enhanced compared to the $O(v_F^4)$ order up
  of $10\%$ for values $v_F\le 0.4$ characteristic to baryons in
  compact stars.

\end{abstract}

\maketitle

\section{Introduction}
\label{sec:1}

The pair-breaking neutrino bremsstrahlung is a important process
contributing to the neutrino cooling of a compact star. The rate of
these processes was computed initially at
one-loop~\cite{Flowers:1976ux, Voskresensky:1987hm, Kaminker:1999ez}
and, more recently, including vertex
corrections~\cite{Leinson:2006,Sedrakian:2006ys,Kolomeitsev:2008mc,
  Steiner:2008qz}.  Within the Standard Model the neutrino emission
occurs via vector and axial-vector current interactions.  The one-loop
calculations suggest that the neutrino emission via neutral
vector currents is large compared to the emission via axial vector
currents.  However, the vertex corrections substantially suppress the
emission via vector currents~\cite{Leinson:2006,Sedrakian:2006ys,
  Kolomeitsev:2008mc,Steiner:2008qz} while they leave the axial vector
emission unaffected.  As a result, the neutrino emission via vector
currents turns out to be subdominant compared to the same emission
mediated by the axial-vector current interactions. Vertex corrections
can be important in the context of neutrino scattering in
proto-neutron-star matter, if the superfluidity sets in before the matter becomes
transparent to neutrinos~\cite{Kundu:2004mz}.
    
In this work we compute higher order corrections to the leading order
(vanishing) contribution to the vector current neutrino emission rate.
We adopt the same approach as in Ref.~\cite{Sedrakian:2006ys}, where a
non-zero temperature propagator formalism was used to compute the
vertex corrections to the weak interaction rates in pair-correlated
baryonic matter.  The convenient techniques of Nambu-Gor'kov
propagators were used for a re-summation of particle-hole ladder
diagrams - an approach first developed by Abrikosov and Gor'kov in
electrodynamics of superconductors~\cite{AG} (see also
Ref.~\cite{Abrikosov:1962}).  In this theory the response of
superconductors to external probes is expressed in the language of
propagators at non-zero temperature and density with contact
interactions that do not distinguish among the particle-hole and
particle-particle channels. It is equivalent to the theories initially
advanced by Bogolyubov, Anderson and others, which are based on the
equations of motions for second-quantized operators.  A more involved
approach was developed subsequently on the basis of the ideas of
Fermi-liquid theory for superconductors/superfluids by
Refs.~\cite{Migdal:1967,Leggett:1966zz}. The latter method implements
wave-function-renormalization of quasiparticle spectrum and  higher-order
harmonics in the interaction channels, and postulates particle-hole
(ph) and particle-particle (pp) interactions with different
strength and/or sign. Application of this method in the context of
neutrino emission and collective excitations can be found in
Refs.~\cite{Kolomeitsev:2008mc,Baldo:2011nc}.  In the
preceding paper~\cite{Sedrakian:2006ys} the driving terms in both
channels were taken to be identical (i.e. $v_{pp} = v_{ph}$) and equal to
the lowest order Landau parameter in baryonic matter. This
identification for $v_{ph}$ is consistent with the theory of normal
Fermi-liquids and guarantees the correct limiting form of the response
function in the unpaired state. The same identification for $v_{pp}$
presumes pairing interaction which includes polarization of the medium in
a manner that has been frequently used in the computations of neutron
and neutron-star matter~\cite{Sedrakian:2006xm}.

It is now well established that the pair-breaking process vanishes to
the leading order (LO) in small momentum transfer for both neutral
vector and axial-vector currents ~\cite{Flowers:1976ux,
  Voskresensky:1987hm, Kaminker:1999ez,Leinson:2006,Sedrakian:2006ys,
  Kolomeitsev:2008mc,Steiner:2008qz}.  The purpose of this work is to
revise and extend the results of Ref.~\cite{Sedrakian:2006ys}
concerning the neutrino emission via neutral vector currents.  First,
we recompute and correct the next-to-leading (NLO)
contribution. Second, we extend the calculation to the
next-to-next-to-leading order (NNLO) contribution and access the
convergence of the series.

The main focus and motivation of this work is the computation of the
neutrino emissivity of baryonic matter in compact
stars~\cite{Shapiro,Sedrakian:2006mq}.  The low-energy neutral weak
current interaction Lagrangian describing the interaction of neutrino
field $\psi$ and baryonic current $j_{\mu}$ is given by \be {\cal L}_W
= -\frac{G_F}{2\sqrt{2}} j_{\mu}\bar\psi \gamma^{\mu}(1-\gamma^5)\psi,
\ee where $G_F$ is the Fermi coupling constant. The current of baryons
for each $B$-baryon is \be j_{\mu} = \bar \psi_B
\gamma_{\mu}(c^{(B)}_V-c^{(B)}_A\gamma^5) \psi_B , \ee where $\psi_B$
are the quantum fields of the baryons and $c^{(B)}_V$ and $c^{(B)}_A$
are the vector and axial vector couplings, respectively.

The rate at which neutrinos are radiated from matter (neutrino
emissivity) is given by (for derivations using equilibrium or
transport techniques 
see Refs.~\cite{Voskresensky:1987hm,Sedrakian:1999jh})
\bea
\label{eq:emiss1} 
\varepsilon_{\nu\anu}&=& - 2\left(
  \frac{G_F}{2\sqrt{2}}\right)^2\int\!  d^4q g(\omega)\omega
\sum_{i=1,2}\int\!\frac{d^3q_i}{(2\pi)^32 \omega_i} \nonumber\\
&\times&\Img[
L^{\mu\lambda}(q_i)\,\Pi_{\mu\lambda}(q)]\delta^{(4)}(q-\sum_iq_i),
\eea where $q_i = (\omega_i,\vecq_i)$, $i=1,2$ are the neutrino
momenta, $g(\omega) = [{\rm exp}(\omega/T) - 1]^{-1}$ is the Bose
distribution function, $\Pi_{\mu\lambda}(q)$ is the retarded
polarization tensor of baryons, and \bea L^{\mu\nu}(q_1,q_2) &=& 4
\Big[q_1^{\mu}q_2^{\nu}+q_2^{\mu}q_1^{\nu}- (q_1\cdot q_2)g^{\mu\nu}
\nonumber\\
&&\hspace{3cm}-i\epsilon^{\alpha\beta\mu\nu}q_{1\alpha}q_{2\beta}\Big]
\eea 
is the leptonic trace.  Here the emissivity is defined per
neutrino flavor, i.e.,  the full rate of neutrino radiation through weak
neutral currents is larger by a factor $N_f$ - the number of
neutrino flavors. (We will assume $N_f= 3$ massless neutrino
flavors.)

The paper is organized as follows.  In Sec.~\ref{sec:2} we set the
stage and introduce the correlation functions needed for a description
of the superfluid baryonic matter.  In Sec.~\ref{sec:3} we discuss the
vertex functions and polarization tensors for neutral vector current
interactions of baryons and neutrinos and their double expansion in
small parameters of the theory. Section~\ref{sec:4} computes the
neutrino emissivity via neutral vector currents up to NNLO order.
Section~\ref{sec:5} contains a brief discussion of the results. Some
details of calculations are relegated to Appendix~\ref{AppendixA} and
the differences to Ref.~\cite{Sedrakian:2006ys} are discussed in 
Appendix~\ref{AppendixB}.


\section{Propagators}
\label{sec:2}

Consider low-density baryonic matter with attractive interaction in
the $^1S_0$-channel. The interaction Lagrangian for the baryons is 
\bea \label{eq:lagrangian_pp} {{\cal L}}_{int} &=&
-v_{pp}\sum_{\vecp_1\neq \vecp_1'}
\psi_{B,\uparrow}^{\dagger}(\vecp'_1)\psi_{B,\downarrow}^{\dagger}(\vecp'_2)
\psi_{B,\downarrow}(\vecp_2)\psi_{B,\uparrow}(\vecp_1)\nonumber\\
&\times&\delta(\vecp_1+\vecp_2-\vecp_1'-\vecp_2'), \eea where $v_{pp}<0$ is
the four-point coupling responsible for binding the particles in
Cooper pairs.  The imaginary-time momentum-space correlators are given
by the $2 \times 2$ Nambu-Gor'kov matrix
\bea \label{eq:propmatrix} {\cal G}_{\sigma,\sigma'}(i\omega_n,\vecp)
&=& \left(\begin{array}{cc} \hat G_{\sigma\sigma'}(i\omega_n,\vecp)
    &\hat
    F_{\sigma\sigma'}(i\omega_n,\vecp)   \\
    \hat F^+_{\sigma\sigma'}(i\omega_n,\vecp) & \hat
    G^+_{\sigma\sigma'}(i\omega_n,\vecp)
\end{array}
\right).
\eea
The elements of the matrix are time-order  correlators of the baryon
field $\psi_B$ and $\psi^{\dagger}_B$; in the frequency-momentum
domain these are given by [Ref. \cite{Sedrakian:2006ys}, Eqs. (5) and (6)]
\bea \label{Prop_P}
\hat G_{\sigma\sigma'}(i\omega_n,\vecp) 
&=& \delta_{\sigma\sigma'}\left(
\frac{u_p^2}{i\omega_n-\ep_p} +\frac{v_p^2}{i\omega_n+\ep_p} \right),\\
    \label{Prop_F}
    \hat F_{\sigma\sigma'}(i\omega_n,\vecp) &=& - i\sigma_y
    u_pv_p\left(\frac{1}{i\omega_n-\ep_p}-\frac{1}{i\omega_n+\ep_p}\right),
    \eea 
where $ F_{\sigma\sigma'}^{+}(i\omega_n,\vecp) =
    F_{\sigma\sigma'}(i\omega_n,\vecp)$, $\omega_n = (2n+1)\pi T$ is
    the fermionic Matsubara frequency, $\sigma_y$ is the $y$ component
    of the Pauli-matrix in spin space, $ u_p^2
    =(1/2)\left(1+{\xi_p}/{\ep_p}\right) $ and $v_p^2 = 1-u_p^2$ are
    the Bogolyubov amplitudes, $\xi_p$ is the single particle energy
 in the unpaired state
    corresponding to the momentum $p$, and $\ep_p =
    \sqrt{\xi_p^2+\Delta_p^2}$ is the single particle energy in the
    paired state, with $\Delta_p$ being the (generally momentum- and
    energy-dependent) gap in the quasiparticle spectrum.  The
    propagator for the holes is defined as $\hat G_{\sigma\sigma'}^{+}
    (i\omega_n,\vecp) = \hat G_{\sigma\sigma'}(-i\omega_n,-\vecp) $.
    For an $S$-wave condensate the spin structure of the propagators
    can be made explicit by writing $\hat
    G_{\sigma\sigma'}(i\omega_n,\vecp)=\delta_{\sigma\sigma'}
    G(i\omega_n,\vecp)$, $\hat F_{\sigma\sigma'}(i\omega_n,\vecp) = -
    i\sigma_yF(i\omega_n,\vecp)$, {\it etc}.

   The single-particle energies in the superconducting state, $\ep_p$,
   are the solution of the Dyson-Schwinger equation $ {\cal G}_{\sigma
     ,\sigma'}^{-1} = {\cal G}_{\sigma ,\sigma', 0}
   ^{-1}-\Sigma_{\sigma ,\sigma'}, $ where ${\cal G}_{\sigma,\sigma',
     0}$ is the diagonal free-particle propagator, whereas the
   self-energy $\Sigma_{\sigma ,\sigma'}$ is a $2\times 2$ matrix
   analogous to Eq.~(\ref{eq:propmatrix}).  The diagonal elements of
   the self-energy matrix renormalize the quasiparticle spectrum and
   the density of states on the Fermi surface in the normal state. We
   will assume that such a program has been carried out and write the
   particle energy in the unpaired state as $\xi_p = v_F (p-p_F)$,
   where $v_F$ and $p_F$ are the (effective) Fermi velocity and
   momentum.  For contact pairing interactions
   (\ref{eq:lagrangian_pp}) the gap function (the off-diagonal 
   self-energy) is momentum-frequency independent; hereafter we
   set  $\Delta_p \equiv \Delta$.

\begin{figure*}[t]
\begin{center}
\includegraphics[height=5.5cm,width=14cm]{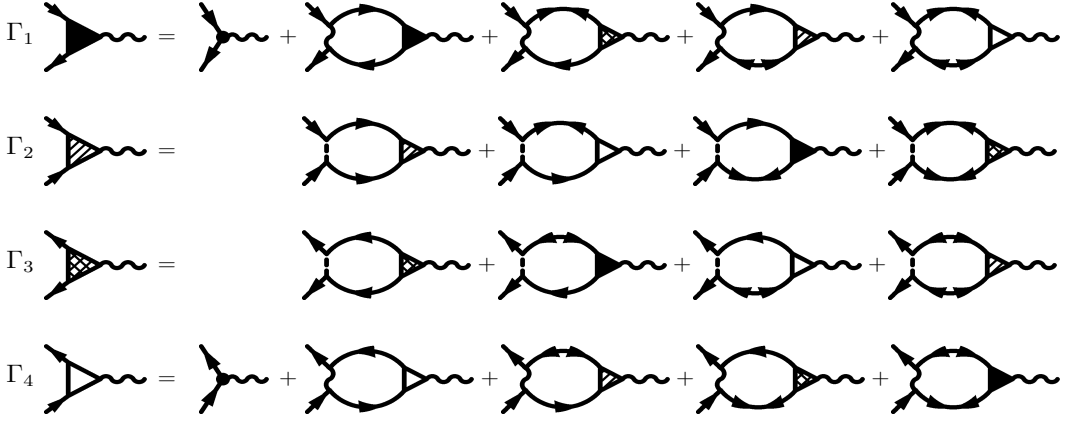}
\end{center}
\caption[] { A diagrammatic representation of the coupled integral
equations for the effective weak vertices in superfluid baryonic
matter.  The ``normal'' propagators for particles (holes) are shown
by single-arrowed lines directed from left to right (right to left).
  The double arrowed lines correspond to the ``anomalous''
propagators $F$ (two incoming arrows) and $F^+$ (two outgoing
arrows).  The ``normal'' vertices $\Gamma_1$ and $\Gamma_4$ are
shown by full and empty triangles.  The ``anomalous'' vertices
$\Gamma_2$ and $\Gamma_3$ are shown by hatched and shaded
triangles. The horizontal wavy lines represent the low-energy
propagator of $Z^0$ gauge boson. The vertical dashed lines stand for
the particle-particle interaction $v_{pp}$; wavy lines represent
particle-hole interaction $v_{ph}$.  }
\label{fig1}
\end{figure*}
\section{Effective vertices and polarization tensors for weak interactions }
\label{sec:3}

\subsection{Vertex functions}

This section reviews and complements the discussion of the
renormalization of the weak vertices in Ref. ~\cite{Sedrakian:2006ys}.
The diagrammatic approach and the mathematical structure of the theory
is essentially the one found in
Refs.~\cite{AG,Abrikosov:1962,Migdal:1967}, but its physical content
(the electroweak dynamics of baryons) is distinct.  We start by
stating the equations that are obeyed by the weak vector-current
vertex $\Gamma_{\mu}=(\Gamma_0,\vec\Gamma)$ in pair-correlated matter.
We use the convention in which greek indices run through temporal 0
and spatial 1,2,3 components, i.e., $\mu=0,1,2,3$.  The latin indices
run through spatial components only.

The temporal ($\mu = 0$) component of the vector-current vertex has
four components in the Nambu-Gor'kov space in general. These
components are given by the following integral equations  [Ref. \cite{Sedrakian:2006ys}, Eqs. (20)-(22)]
\bea
\label{g1}
\hat\Gamma_1 &=& \Gamma_0 
             +  v_{ph} (G\Gamma_1G
             +       \hat F\hat\Gamma_3G
             +       G\hat\Gamma_2\hat F
             +       \hat F\Gamma_4\hat F),  \\ 
\label{g2}
\hat\Gamma_2 &=&  \quad \quad v_{pp} (G\hat\Gamma_2G^{+}
             +       \hat F\Gamma_4G^{+}
             +       G\Gamma_1\hat F
             +       \hat F\hat\Gamma_3\hat F), \\ 
\label{g3}
\hat\Gamma_3 &=& \quad \quad v_{pp} (G^{+}\hat\Gamma_3G + \hat
F\Gamma_1G + G^{+}\Gamma_4\hat F
+       \hat F\hat\Gamma_2\hat F),\\
\label{g4}
\hat\Gamma_4 &=& \Gamma_0 + v_{ph} (G^{+}\Gamma_4G^{+} + \hat
F\Gamma_1 \hat F + \hat F \hat\Gamma_2 G^{+} + G^{+}\hat\Gamma_3\hat
F), \nonumber\\
\eea 
which are written in the operator form and are shown diagrammatically
in Fig.~\ref{fig1}.  The $\mu=0$ index on vertices is suppressed, the
bare vertex is $\Gamma_0 = 1$, and the hats over the vertices indicate
that these are $2\times 2$ matrices in the spin-space, specifically
$\hat \Gamma^{(2)}= -i\sigma_y \Gamma^{(2)}$ and  $\hat \Gamma^{(3)}=
-i\sigma_y \Gamma^{(3)}$.

The vertices $\Gamma_1\dots \Gamma_4$ are functions of the four-momentum
transfer $q$ and are independent of the incoming and outgoing
momenta.  Therefore, each term on the
right hand side of Eqs.~(\ref{g1})-(\ref{g4}) contains a polarization insertion
defined as 
 [Ref. \cite{Sedrakian:2006ys}, Eq. (25)]
\bea \label{eq:pi1}
\Pi_{XX'}(q) &=& T\int\frac{d^3\vecp}{(2\pi)^3}\sum_{ip_n} X(p)X'(p+q) \nonumber\\
&=& \frac{\nu(0)}{4}\int_{-1}^{1}\! dx (X * X'), 
\eea 
where $T$ is the temperature, 
 $X,X'\in \{G,F,G^+,F^+\}$, $p\equiv (ip_n,\vecp)$ with $p_n$ being the
fermionic Matsubara frequency, $\nu(0) = m^* p_F/\pi^2$ is the density
of states at the Fermi surface, and $x$ is the cosine of the angle formed
by the vectors $\vecq$ and $\vecp$. The convolution (or loop) is
defined in Eq.~\eqref{eq:pi1} as
\be\label{convolution}
(X * X')=T\int_{-\infty}^{\infty}d\xi_p \sum_{ip_n} X(p)X'(p+q).
\ee 
Taking the integration limits over the 
infinite  range $-\infty\le \xi_p\le \infty$ is a valid 
approximation in the weakly coupled superconductivity. 
One may now exploit relations among the loops
[Ref. \cite{Sedrakian:2006ys}, Equations (26) and (A8)]: $(G* F) \simeq
-(F* G),$ $(G^{+}* G) \simeq (G* G^{+})$, and $(F* G^{+}) \simeq
-(G^{+}* F)$ to show that Eqs.~(\ref{g2}) and (\ref{g3}) have a
solution only if $\Gamma_2+\Gamma_3 =0$, i. e., $\Gamma_2=-\Gamma_3 $.
Eqs. (\ref{g1}) and (\ref{g4}) transform into each other on reversal
of the time direction, which implies that $ \Gamma_1(\omega,\vecq) =
{\cal T}\Gamma_4(\omega,\vecq)$, where ${\cal T}$ is an operator
affecting the reversal. It is $+1$ for the scalar vertex and $-1$ for
any vector vertex. This property is most easily checked by a
calculation of the loops that change their sign, e. g. $(F*F)$, for
scalar and vector vertices. The solution of the linear system of
equations for the temporal part of the vector-current vertex
(\ref{g1})-(\ref{g4}) reads \bea
\label{Gamma1}
\Gamma_1 &=& \frac{\Gamma_0 {\cal C}}{{\cal C}-v_{ph}({\cal
      A}^+{\cal C}-{\cal BD}^+)},\\
\label{Gamma2}
\Gamma_2 &=& -\frac{\Gamma_0 {\cal D}^+}{{\cal C}
-v_{ph}({\cal A}^+{\cal C}-{\cal BD}^+)},
\eea
where ${\cal A}^{\cal T}= \Pi_{GG} -\Pi_{FF}{\cal T}$, ${\cal B} = 2\Pi_{FG}$, ${\cal C} =
-[v_{pp}^{-1}- (\Pi_{GG^+} +\Pi_{FF})]$ and ${\cal D}^{\cal T} =
\Pi_{FG^+}{\cal T}+\Pi_{GF}$. 
For  ${\cal T}=1$ one finds  ${\cal B} =-{\cal D}^+$. Upon setting 
$\Gamma_0=1$ we verify that the solutions (\ref{Gamma1}) and (\ref{Gamma2}) 
coincide with those given in Ref.~\cite{Sedrakian:2006ys}, Eqs. (33)
and (34).
\begin{figure*}[tb]
\begin{center}
\includegraphics[height=1.5cm,width=12cm]{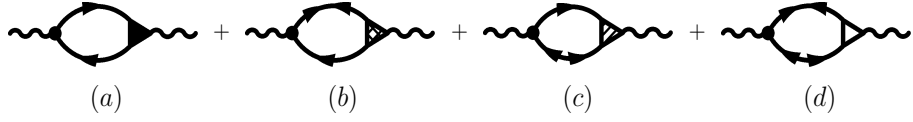}
\end{center}
\caption[] { The sum of polarization tensors contributing to the
vector-current neutrino emission rate.  Note that the diagrams $b$,
$c$, and $d$ are specific to the superfluid systems and vanish in the
unpaired state.  }
\label{fig2}
\end{figure*}
The vector vertex $\vec\Gamma$  is defined in full analogy to the scalar vertex:
\bea\label{gv1}
\vec\Gamma_1 &=& \vec\Gamma_0 
            +  v_{ph} (G\vec\Gamma_1G
            -    F\vec\Gamma_3G
            -    G\vec\Gamma_2 F
            -    F\vec\Gamma_4 F),  \\ 
\label{gv2}
\vec\Gamma_2 &=&  \quad \quad v_{pp} (G\vec\Gamma_2G^{+}
            +       F\vec\Gamma_4G^{+}
            +       G\vec\Gamma_1 F
            -       F\vec\Gamma_3F), \\ 
\label{gv3}
\vec\Gamma_3 &=&  \quad \quad v_{pp} (G^{+}\vec\Gamma_3G
            +        F\vec\Gamma_1G
            +       G^{+}\vec\Gamma_4 F
            -       F\vec\Gamma_2F),
\\
\label{gv4}
\vec\Gamma_4 &=& \vec\Gamma_0 + v_{ph} (G^{+}\vec\Gamma_4G^{+} - F
\vec\Gamma_1 F - F \vec\Gamma_2 G^{+}
-     G^{+}\vec\Gamma_3 F),\nonumber\\
\eea where the same arguments as above imply $\vec \Gamma_2 = -\vec
\Gamma_3$ and $\vec\Gamma_1 = {\cal T} \vec\Gamma_4$.  There are two
vectors available in the problem, therefore the most general
decomposition of any vector vertex in terms of these vectors reads  
\bea \vec\Gamma_{1,2} =
P_{1}\hat p+Q_{1,2}\hat q, \quad \vec\Gamma_0 = P_1 \hat p, 
\eea 
where $\hat p = \vecp/\vert \vecp\vert$, 
$\hat q = \vecq/\vert \vecq\vert$ and
$P_1$ and $Q_{1,2}$ are scalar functions, which depend on the moduli
of these vectors. The solution of the set of linear equations
\eqref{gv1}-\eqref{g4} implies that 
\bea
Q_1 &=& \frac{v_{ph}P_1[\tilde{\cal A}{\cal C}
-{\cal B}\tilde{\cal D}]}
{{\cal C}-v_{ph}[{\cal A}^+{\cal C}-{\cal B}{\cal D}^+]},\\
Q_2 &=& 
-P_1\frac{
\tilde{\cal D} -v_{ph}[{\cal A}^+\tilde {\cal D}
-\tilde{\cal A}{\cal D}^+
]
}{{\cal C}-v_{ph}[{\cal A}^+{\cal C}-{\cal B}{\cal D}^+]
},
\eea
where $\tilde{\cal A} \equiv {\cal A}^-(\hat q\cdot \hat p)$ and 
$\tilde{\cal D} \equiv {\cal D}^-(\hat q\cdot \hat p)$ are the first
moments of the ${\cal A}^-$ and ${\cal D}^-$ integrals with respect 
to the cosine of the angle formed by the vectors $\vecq$ and $\vecp$.

\subsection{Polarization tensors}

The 00 component of the vector polarization tensor is 
given by the sum of four terms
\bea 
\Pi_{00} =\Gamma_0 (G\Gamma_1G + \hat F\hat\Gamma_3G + G\hat\Gamma_2\hat
F + \hat F\Gamma_4\hat F),
\eea 
which are shown diagrammatically in Fig.~\ref{fig2}.
Exploiting the
relations $\Gamma_1 = {\cal T}\Gamma_4$ and $\Gamma_2=-\Gamma_3$ one finds
$\Pi_{00} = \Gamma_0({\cal A}^+\Gamma_1 +{\cal B}\Gamma_2).$ Then,
substituting the vertices (\ref{Gamma1}) and (\ref{Gamma2}), 
we obtain
\bea \label{eq:pi00}
\Pi_{00} = \frac{\Gamma_0 ({\cal A}^+{\cal C}-{\cal B} {\cal
    D}^+) }{{\cal C}-v_{ph}({\cal A}^+{\cal C}-{\cal BD}^+)}.  
\eea
Because ${\cal B} = -{\cal D}^+$ for ${\cal T}=1$ one recovers the
polarization tensor given by Eq. (35) of
Ref.~\cite{Sedrakian:2006ys}. For 
the spatial components of the vector current 
polarization tensor one finds 
\bea \label{eq:PI_vector}
\Pi_{ij} &=& \hat p_i v_F^2{\cal A}^-\hat p_j+
\hat p_i v_F^2\frac{v_{ph} [\tilde{\cal A}{\cal C} -{\cal
     B}\tilde{\cal D}]} {{\cal C}-v_{ph}[{\cal A}^+{\cal C}-{\cal
     B}{\cal D}^+]}{\cal A}^+\hat q_j \nonumber\\ 
&-&\hat p_i v_F^2{\cal B}\frac{\tilde{\cal D} -v_{ph}[{\cal A}^+\tilde {\cal D}
   +\tilde{\cal A}{\cal D}^+]}{{\cal C}-v_{ph}[{\cal A}^+{\cal
     C}-{\cal B}{\cal D}^+]
 }\hat q_j,
\eea 
where we approximated the value of the $P_1$  by  $v_F$.

\subsection{Expanding the polarization tensor}

We will use a double-expansion in small parameters which arise in the
limit $\vert\vecq\vert\to 0$ with other parameters held fixed.  Contrary to
Ref.~\cite{Sedrakian:2006ys}, we use the ``symmetric'' kinematics,
where perturbed quasiparticle energies in the normal and
superconducting state are written, respectively, as $\xi_{\pm} =\xi_p
\pm (\vecv \cdot \vecq)/2$ and $\ep_{\pm} = \sqrt{\xi_{\pm}^2+\Delta^2}$.
The loops ${\cal A},\, {\cal B},\, {\cal C}$ and $ {\cal D}$ are
products of the convolution $(F*F)$ and prefactors which depend only
on the four-momentum transfer [Ref. \cite{Leggett:1966zz}, Eqs. (15)
and (17)]. These pre-factors are expanded in the small parameter
$\delta = \vert q \vert v_F/\omega$. The expansion of the convolution
$(F*F)$ around $\xi_p \simeq \xi_- \simeq \xi_+$ is carried
out with respect to the second small paramater $\eta = v_F \vert 
\vecq\vert /\xi_p$; for details see  Appendix~\ref{AppendixA}.
Specifically, we write the Taylor expansion as 
\be \label{eq:Lexp} 
(F*F^+)=\sum_n(F*F^+)_n  \delta^n x^n ,
\ee 
where $x \equiv \hat q\cdot \hat p$, and the factor $\eta^n/\delta^n$ is
absorbed in the definition of the coefficients $(F*F^+)_n$ of the Taylor
expansion \eqref{eq:Lexp}.  
Substituting the expressions for the functions ${\cal
  A}^+,\, {\cal B},\, {\cal C}$, and $ {\cal D}^+$ in
Eq.~(\ref{eq:pi00}) we obtain the scalar polarization tensor
truncated at order $O(\eta^6)$ and $O(\delta^6)$:
\bea \frac{\Pi_{00}  (q)}{2\nu(0)} &=& \overline{\left[
   -\frac{x\delta}{1-x\delta}-1\right]{(F*F^+)}}
\nonumber\\
&+&\frac{\overline{(F*F^+)}}{\overline{(1-x^2\delta^2)(F*F^+)}}\ 
\overline{(1+x\delta)
 (F*F^+)}\nonumber\\
&\simeq& -\frac{4 \delta ^4}{45} \left(1+\frac{25 \delta
   ^2}{21}\right)(F*F^+)_0 - \frac{4\delta ^6}{45} (F*F^+)_2\nonumber\\
&+&O(\delta^8),
\eea  
where $\overline{(\dots)} \equiv (1/2)\int_{-1}^1 \, dx(\dots) $. 
Similar expansion for the spatial part of the vector current
polarization tensor \eqref{eq:PI_vector} gives 
\bea \frac{ \Pi_{ii} (q)}{2v_F^2\nu(0)} &=&
\overline{\left[ -\frac{x\delta}{1-x\delta}\right](F*F^+)}\nonumber\\
&+&\frac{\overline{(x+x^2\delta)(F*F^+)}}
{\overline{(1-x^2\delta^2)(F*F^+)}}~ \overline{x^2\delta (F*F^+)}
\nonumber\\
&\simeq&-\left(\frac{2\delta^2}{9} +\frac{22\delta^4}{135}
  +\frac{74\delta^6}{567}\right) (F*F^+)_0\nonumber\\
&-&\left(\frac{14\delta^4}{135} +\frac{286\delta^6}{2835}\right)
 (F*F^+)_2 +O(\delta^8).  \nonumber\\
\eea 
In this last expression we need to keep only the terms up to
$\delta^4$, because the vector vertices are proportional to the Fermi
velocity, i. e.,  a small parameter.

\section{Vector current neutrino emissivity}
\label{sec:4}
\begin{figure}[t]
\begin{center}
\includegraphics[height=8.0cm,width=\linewidth]{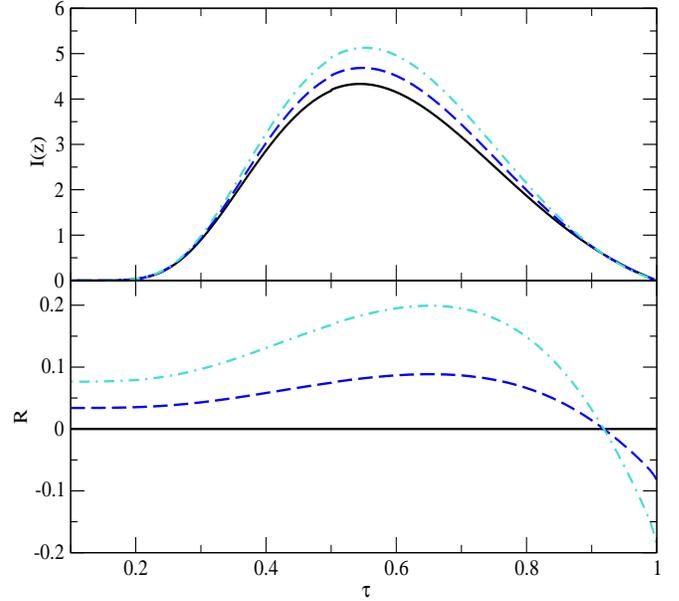}
\end{center}
\caption[] {(Color online) {\it Upper panel:} Dependence of the integral \eqref{eq:I}
  on the reduced temperature $\tau$ for  $v_F = 0$ (solid
  line), 0.4 (dashed line) and 0.6 (dashed dotted line).  {\it Lower
    panel}. The ratio $R$ of the integral \eqref{eq:I} computed at
  $O(v_F^6)$ order to its value computed at order $O(v_F^4)$.}
\label{fig3}
\vskip 0.5cm
\end{figure}

To obtain the neutrino pair bremsstrahlung emissivity we 
contract  the baryonic polarization tensor with the trace over
the leptonic currents, see Eq. \eqref{eq:emiss1}.
The result can be cast as 
\be \label{eq:emiss2} \epsilon = \frac{G^2c_V^2
  N_f}{48\pi^4}\int_0^{\infty}d\omega g(\omega) \omega J(\omega), \ee
where  $c_V = 1$
for neutrons and $c_V = 0.08$ for protons, $N_f =3$ is the number of
neutrino flavors in the Standard Model, and 
\bea J(\omega) &=&
\int_0^{\omega} d\vecq \vecq^2 (\vecq^2-\omega^2)\text{Im}
\left[\Pi_{00}(\omega,q) -\Pi_{ii}(\omega,q)\right]\nonumber\\
&=&-\frac{8\omega^5 \nu(0)v_F^4}{405}{\text{Im}(F*F^+)_0} [1+\gamma
v_F^2] , \eea 
where $\vecq$ is the momentum transfer and the
coefficient $\gamma$ is defined by  Eq.~\eqref{gamma} of
Appendix~\ref{AppendixA}.  Substituting this result in
Eq.~(\ref{eq:emiss2}) we find 
\bea
\epsilon &=&\frac{16 G^2 c_V^2 \nu(0) v_F^4 }{1215\pi^3} I(z) T^7 ,
\eea
where $z = \Delta/T$ and
\bea I(z)&=& z^7 \int_{1}^{\infty}\!\!  
\frac{dy\  y^5}{\sqrt{y^2-1}} f \left(z y\right)^2 \left[1+
  \left(\frac{7}{33}+\frac{41}{77}\gamma\right)v_F^2\right].
\nonumber\\
\label{eq:I}
\eea 
To order $v_F^4$ this result agrees with those obtained by
Refs.~\cite{Leinson:2006,Kolomeitsev:2008mc}. The numerical result at
order $O(v_F^6)$ and their comparison to order $O(v_F^4)$ result are
shown in Fig.~\ref{fig3}.  One observes that the corrections to the
emissivity are small, i. e., the series expansion is justified. The
emissivity turns out to be enhanced by about $10\%$ for $v_F = 0.4$
and by about $20\%$ for $v_F = 0.6$ in the temperature range $0\le
\tau \le 0.92 $, where $\tau = T/T_c $ with $T_c$ being the critical
temperature of pairing phase transition. In the close vicinity of $T_c$ the
emissivity is reduced.

\section{Concluding remarks}
\label{sec:5}

We obtained the neutrino emissivity via neutral vector currents in a
perturbative expansion up to order $O(v_F^6)$. The order $O(v_F^4)$
term corrects our previous NLO result~\cite{Sedrakian:2006ys} and is
in agreement with the expressions given in
Refs.~\cite{Leinson:2006,Kolomeitsev:2008mc}.  (The expansion of the
vector polarization tensor in Ref.~\cite{Sedrakian:2006ys} gave a
mathematically spurious NLO contribution that is linear in the nucleon
recoil~\footnote{ Linear in momentum transfer perturbation of
  quasiparticle energy is $ (\vecq \cdot \vecv)\sim qv_F x$ [the
  recoil term is $O(q^2)$]. Any quantity which is expanded to the
  linear order in this perturbation will vanish upon integration over
  $x$ in symmetrical limits. Therefore, truncating at linear order in
  the perturbation Ref.~\cite{Sedrakian:2006ys} do not find terms
  $\propto v_F$. In view of this observation the statement in
  Ref.~\cite{Kolomeitsev:2008mc} that the terms $\propto v_F$ were
  dropped in Ref.~\cite{Sedrakian:2006ys} should not be taken
  literally.}.  As a result, the emissivity was suppressed by a factor
$T/m$, which arises from the recoil term, instead of the correct
factor $v_F^4$).

From a practical point of applications in  astrophysics, the
$O(v_F^4$) order term in the vector current neutrino emissivity is
unimportant. Indeed, it is negligible compared to the axial vector
neutrino emissivity as computed in
Refs.~\cite{Flowers:1976ux,Kaminker:1999ez, Kolomeitsev:2008mc}, which
remains unaffected by the vertex corrections. Nevertheless, the
present reassessment removes the disagreement between 
Refs.~\cite{Leinson:2006,Kolomeitsev:2008mc,Steiner:2008qz} and 
Ref.~\cite{Sedrakian:2006ys}, and confirms that the nonvanishing
contribution to the vector-current emissivity arises at order
$O(v_F^4)$. Similarly, it revises the results of
Ref.~\cite{Sedrakian:2010xe} concerning the density response functions
of cold neutron matter.
However, we note that the
absence of $O(v_F^2)$ terms in the vector response is not protected by
any symmetry of the theory (i.e., conservation law), whereas the $O(1)$
contribution remaining in one-loop calculations are prohibited by
baryon number conservation.

The computation of the $O(v_F^6)$ order contribution to the
vector-current neutrino emissivity shows that the corrections to the
leading non-zero term are below $10\%$ for values $v_F\le 0.4$
characteristic of baryons in compact stars.  This result provides an
evidence of the convergence of the series expansion of the
vector-current polarization tensor in the regime where the momentum
transfer is small compared to other relevant scales.

\section*{Acknowledgments}
This work was in part supported by the Deutsche Forschungsgemeinschaft
(Grant No. SE 1836/1-2).

\appendix

\begin{widetext}
\section{Expansion of the loop function}
\label{AppendixA}

The purpose of this appendix is to determine the coefficients of the
expansion of the $(F*F)$ loop in powers of the $\eta$ parameter, 
Eq. (\ref{eq:Lexp}).
The Matsubara sum for the  $FF$ loop gives (Eq. (A12) of Ref. \cite{Sedrakian:2006ys})
\bea
\label{A1}
T\sum_{ip} F(ip,\vecp) \dF (ip+iq, \veck) 
&=& u_pu_kv_pv_k \Biggl\{\left[\frac{1}{iq+\ep_p-\ep_k}-\frac{1}{iq-\ep_p+\ep_k}
\right]\left[f(\ep_p)-f(\ep_k)\right]\nonumber\\
&+&  \left[ \frac{1}{iq+\ep_p+\ep_k}- 
\frac{1}{iq-\ep_p-\ep_k}\right]\left[1-f(\ep_p)-f(\ep_k) \right]\Biggr\}.
\eea
 For contact pairing interactions
$u_pu_kv_pv_k = \Delta^2/4\ep_p\ep_k.$
After an analytical continuation with retarded boundary condition,
$i\omega_n\to \omega+i\delta$, we obtain for the convolution derived
from Eq.~\eqref{A1}
\bea
\label{A2}
(F*F^+) &=&
-\frac{\Delta^2}{2}\int d\xi_p\Bigg\{\frac{(\ep_p-\ep_k)}{\ep_p\ep_k}
\left[  \frac{f(\ep_p)-f(\ep_k)}{\omega^2-(\ep_p-\ep_k)^2+i\delta}\right]
+\frac{(\ep_p+\ep_k)}{\ep_p\ep_k}
\left[ \frac{1-f(\ep_p)-f(\ep_k)}{\omega^2-(\ep_p+\ep_k)^2+i\delta}\right]\Bigg\}.
\eea
\end{widetext} 
To obtain the series expansion approximation to the integral on the
right hand side we expand first the integrand in series and carry out the
integrations term by term. In doing so we will need to compute the
imaginary parts of integrals of the type \be I_n =
\int_{-\infty}^{\infty} \! d\xi
\,\,\frac{f_n(\ep)}{(\omega^2-4\ep^2)^n+i\delta}, \ee where at the
order we are working we need only integrals with $n=1,2$.  For $n=1$
the imaginary part of the integral is obtained with the use of the
Dirac identity. Upon the following change of variables $\xi\to \ep =
\sqrt{\xi^2+\Delta^2}, \quad \xi d\xi = \ep d\ep$, $\Delta \le \ep \le
\infty$, the integral acquires a factor of 2. A straightforward
computation gives 
\bea \label{A3} 
 {\rm Im}I_1 &=& {\rm Im}
  ~\left[\int_{\infty}^{\infty} \!\!d\xi~
    \frac{f_1(\ep)}{\omega^2-4\ep^2+i\delta}\right]\nonumber\\
&=&-\frac{\pi}{2}
\frac{\theta(\omega-2\Delta)}{\sqrt{\omega^2-4\Delta^2}}f_1\left(\frac{\omega}{2}\right).
 \eea 
Consider now the case $n=2$, i. e., the integral \be {\rm Im}I_2
={\rm Im} ~\left[\int_{\infty}^{\infty} \!\!d\xi~
  \frac{f_2(\ep)}{(\omega^2-4\ep^2)^2+i\delta}\right].  \ee We first
change the variable $\xi \to \ep = \sqrt{\xi^2+\Delta^2}, \quad \xi
d\xi = {\ep d\ep}$ and rewrite the integral as \bea {\rm Im}I_2
&=&2{\rm Im} ~\left[\int_{\Delta}^{\infty} \!\!  \frac{\ep
    d\ep}{\sqrt{\ep^2-\Delta^2}}
  \frac{f_2(\ep)}{(\omega^2-4\ep^2)^2+i\delta}\right]. 
\eea 
It is convenient to carry out a second transformation of variables
defined  as $ z =
4\ep^2, \quad dz = 8\ep d\ep, \quad 4\Delta^2\le z\le \infty.  $
Implementing the transformation we obtain \bea {\rm Im}I_2
&=&2\pi\left[\int_{4\Delta^2}^{\infty} \!\!d
  z~\frac{1}{4\sqrt{z-4\Delta^2}}
  f_2(\sqrt{z}/2)\delta^{^{(1)}}(z-\omega^2)\right], \nonumber\\
\eea 
where we used the formula 
\be \frac{1}{D^{k+1}+i\delta} =
\frac{P}{D^{k+1}}-i\pi\frac{(-1)^k}{k!}\delta^{(k)}(D) \ee where $P$
denotes the principal value and $\delta^{(k)}(D)$ denotes the $k$-th
derivative of the delta function. In the case $k=1$ we obtain \be
\frac{1}{D^2+i\delta} = \frac{P}{D^{2}}+i\pi\delta^{(1)}(D).  \ee We
further use the formula \be \int f(x)\delta'(x-a) = - \int
f^{'}(x)\delta(x-a), \ee to obtain the final result
\bea \label{eq:genint2} 
 {\rm Im}I_2 &=&{\rm Im}
  \left[\int_{\infty}^{\infty} \!\!d\xi~
    \frac{f_2(\ep)}{(\omega^2-4\ep^2)^2+i\delta}\right]
\nonumber\\
&=&-\frac{\pi}{2} \theta(\omega-2\Delta)
  \frac{d}{dz}\left[\frac{f_2(\sqrt{z}/2)}{\sqrt{z-4\Delta^2}}
  \right]_{z = \omega^2}.
\eea 
To carry out the expansion we write the variables as 
\be
\label{eq:xi}
\xi_{\pm} = \xi (1\pm \eta x),\quad \eta = \frac{qv_F}{\xi} ,  
\ee 
and substitute \eqref{eq:xi} into the spectra $\ep_{\pm} =
\sqrt{\xi_{\pm}^2+\Delta^2}$. Expanding the kernel to leading order we obtain
\bea
(F*F^+)_0 &=& \Delta^2 \int_{-\infty}^{\infty} d\xi
\frac{\tanh (\ep/2T)}{\ep (\omega^2-4\ep^2)}.
\eea
Using Eq.~\eqref{A3} one finds
\bea
\text{Im}(F*F^+)_0
& = &-\frac{\pi\Delta^2}{\omega}
\frac{\theta(\omega-2\Delta)}{\sqrt{\omega^2-4\Delta^2}} \tanh
\left(\frac{\omega}{4T}\right), 
\eea 
where $\ep = \sqrt{\xi^2+\Delta^2}$.
The coefficient of the second order term is given by
\bea 
\text{Im} (F*F^+)_2 =
-\frac{\pi\Delta^2}{\omega}
\frac{\theta(\omega-2\Delta)}{\sqrt{\omega^2-4\Delta^2}} 
[h_1\left(\omega\right)
+h_2\left(\omega\right)], \nonumber
\eea
where the
contributions arising from integrals wit $n=1$ and 2, respectively, are
\bea h_1&\equiv& \frac{1}{4y^2}\text{sech}^2\left(\frac{\omega }{4 \ T}\right)
\Bigg[\frac{\omega}{2T}+\left(2 y^2-3\right) \sinh \left(\frac{\omega
      \ }{2 T}\right) \nonumber\\
&-& \frac{\omega ^2}{4T^2} \left(y^2-1\right) \tanh
  \  \left(\frac{\omega }{4 T}\right)\Bigg],\\
h_2&=& \frac{1}{4y^2} \text{sech}^2\left(\frac{\omega }{4
    T}\right)\left[ {\frac{\omega}{2T} \ - \frac{\left(2
        y^2-3\right)}{y^2-1} \sinh \left(\frac{\omega }{2
        T}\right)}\right].  \nonumber\\
\eea 
Note that the function $h_{1,2}$ contain a factor $\eta^2/\delta^2
=(4y^2)/(y^2-1)$, with $ y = \Delta/2\omega$. The ratio appearing in
Eq. \eqref{eq:I} is then given by
\bea \label{gamma}
\gamma &=& 
\frac{\text{Im}\, (F*F^+)_2}{\text{Im}\,(F*F^+)_0}
= \frac{1}{4y^2}
\text{sech}\left(\frac{\omega }{4 T}\right)
\text{csch}\left(\frac{\omega }{4 T}\right)\nonumber\\
&\times& \Bigg[{\frac{\omega}{T} +
    \frac{(y^2-2)}{y^2-1} \left(2 y^2-3\right) \sinh
    \left(\frac{\omega }{2 T}\right)}\nonumber\\
&-&\frac{\omega ^2}{4T^2}
  \left(y^2-1\right) \tanh \ \left(\frac{\omega }{4 T}\right) \Bigg].
\eea 
Note that the second term in the brackets changes the singular asymptotics
of the integrand in Eq.~\eqref{eq:I} at the end point $y=1$ from
$(y-1)^{-1/2}$ to $(y-1)^{-3/2}$.

\section{Comparison with Ref.~\cite{Sedrakian:2006ys}}
\label{AppendixB}

In this appendix the equations from 
Ref.~\cite{Sedrakian:2006ys} are preceded by roman I.  As already mentioned above
the polarization tensor (\ref{eq:pi00}) coincides with Eq. (I.35).
We now proceed to examine the three loops appearing in these
equations.  Substituting Eqs. (I.A1) and (I.A3) into the expression for
the ${\cal A}(q)$ loop and using the definitions of the Bogolyubov
amplitudes appearing after Eq.~(\ref{Prop_P}) we obtain
 \be\label{B1} {\cal A}(q) = \int\!\!\!
\frac{d^3p}{(2\pi)^3} \left[(\ep+\ep') (\ep\ep'-\xi\xi'+\Delta^2)
  +\omega (\xi'\ep-\xi\ep') \right]\mathscr{G} 
\ee
 with the short-hand
notations $\xi = \xi_{\vecp}$, $\xi' = \xi_{\vecp+\vecq}$ (asymmetrical kinematics)
or $\xi = \xi_{\vecp-\vecq/2}$, $\xi' = \xi_{\vecp+\vecq/2}$ (symmetrical kinematics),
$\ep = \sqrt{\xi^2+\Delta^2}$, $\ep' = \sqrt{(\xi')^2+\Delta^2}$ 
and~\footnote{The second term in Eq. (32) of
 Ref.~\cite{Sedrakian:2006ys} is a typographical error and should be
 dropped. It does not, however,  affect the discussion or the result.}
\be \label{B2} \mathscr{G} = \frac{1}{2\ep\ep'}
\left[\frac{1-f(\ep)-f(\ep')}{\omega^2-(\ep+\ep')^2}\right].  
\ee 
On the other hand, substituting Eqs.~(I.31) and (I.32)  in
Eq.~(I.27), we see that the resulting expression does not contain the
term $\omega (\xi'\ep-\xi\ep')$ in Eq.~(\ref{B1}), which vanishes in the limit 
$\vecq \to 0$. Using Eq.~(I.A2) we further obtain 
\be
\label{B3} {\cal B}(q) = 2\Delta \int\!\!\!\frac{d^3p}{(2\pi)^3}
\left[\omega \ep' + (\ep'+\ep)\xi' \right]\mathscr{G}.  
\ee
Substituting Eq.~(I.31) in Eq.~(I.28) we see that the term $
(\ep'+\ep)\xi' $ in Eq.~(\ref{B3}) is missing. Finally,
using (I.A3) and (I.A4) in the expression for the ${\cal C}$ loop 
we find 
\bea\label{B4} 
 {\cal C}(q) &=& \int\!\!\!\frac{d^3p}{(2\pi)^3}
\Biggl\{ \frac{1-2f(\ep)}{2\ep}
- \Bigl[(\ep+\ep')(\ep\ep'+\xi\xi'+\Delta^2) \nonumber\\
&-& \omega (\xi\ep'+\xi'\ep)
\Bigr]\mathscr{G}\Biggr\}
\eea
and we see that the term $-\omega (\xi\ep'+\xi'\ep)$ is missing in
Eq.~(I.29). Both terms that were dropped vanish in the limit $\vecq \to 0$,
because they are odd in $\xi$ while their convolutions involve
integrations over symmetrical in $\xi$ limits [see Eqs.~(\ref{eq:pi1})
and Eqs.~(\ref{convolution})]. Thus, we conclude that, while the
$\vecq\to 0$ limit of the ${\cal A}$, ${\cal B}$, and ${\cal C}$ loops
were correctly identified, their small $q$ expansion was carried out
in Ref.~\cite{Sedrakian:2006ys} starting from incomplete expressions.

\end{document}